\documentclass[jpcm]{iopart}

\usepackage{epsf,citesort}
\usepackage{graphicx,citesort,amssymb}

\font\eightrm=cmr8

\input{epsf.sty}
\newcommand{\centfig}[2]
{\begin{center}
\leavevmode
\epsfxsize=#1
\epsffile{#2}
\end{center}}

\begin{document}

\title{{\large Soft Condensed Matter (Materia Condensata Soffice)}}
\author{M. E. Cates}

\address{School of Physics,
University of Edinburgh,
\\ JCMB King's Buildings, Mayfield Road\\
Edinburgh EH9 3JZ, Scotland}

\begin{abstract}
This article was written for a multi-volume History of Science, and published in Italian translation: Storia della Scienza, Volume IX, La Grande Scienza, pp 645--656 (Published by Istituto della Enciclopedia Italiana, Fondata da Giovanni Treccani, 2003). In it, I describe the evolution of soft matter physics as a discipline during the 20th century.
\end{abstract}
\maketitle

\section*{Introduction} 
Soft condensed matter physics concerns colloids, polymer solutions, emulsions, foams, surfactant solutions, powders and similar materials. (Domestic examples are respectively paint, engine oil, mayonnaise, shaving cream, talc, etc..)
In each case, the precise molecular composition of the system has only limited influence on the physical behaviour, which is controlled by structure on the mesoscopic scale (between, say, one nanometre and one micron) that is easily reorganized by external influences such as mechanical stress. The unusual mechanical properties of such soft condensed phases are widely exploited both by nature and by mankind, either directly or during processing. An example of the latter is the moulding of a hard plastic object by casting it from the melt: polymer melts represent soft condensed matter `par excellence'. 

\section*{Intellectual and Social Context}
The second half of the 20th century, and especially its final two decades, saw a very strong growth in our understanding of soft condensed matter physics. Several reasons for this can be identified. 

Firstly, there have been a number of industrial drivers for research in this field. Indeed, the wide availability of well-controlled synthetic samples of polymeric and colloidal materials dates back only to the 1940s and 1950s, and can be viewed as a by-product of the emergence, at about that time, of the plastics industry. Before this, studies were reliant on naturally occurring materials (such as natural rubber) and/or ones with relatively poor reproducibility. From the 1950s onwards, the industrial motivation for an improved understanding of the physical properties of polymers was clear. There were several other industrial drivers, such as the increasing transfer (from the 1960s onwards, at least in western countries) of food preparation duties from the kitchen to the factory. This required a far more scientific approach to understanding the mechanical stability of soft materials. Subsequently, mounting environmental and health concerns have maintained this pressure. For example, paints and coatings which used to contain harmful organic solvents are now mainly water-based; many foods that used to contain chemical preservatives or artificially hydrogenated fats no longer do so. In many such cases, it is the better application of physics (the control of structure by  mechanical or thermal processing) that allows less reliance on chemical methods of stabilizing a product.

Setting aside these industrial motivations, soft condensed matter has benefited greatly from broader scientific developments. Much progress has stemmed from improved experimental methods, particularly the use of scattering by laser light, X-rays and neutrons as probes of static and dynamic structure. Laser light scattering (which stemmed from the invention of the laser itself in the late 1950s) has proven particularly valuable because the length scales probed by it are suitable for exploring mesoscopic structure; the dynamic version, in which the object of study is the time correlation of the scattered intensity, allows detailed characterisation of motion on the micron scale. From the 1980s onwards, the power of such methods was extended further through the development of techniques to allow interpretation of laser light scattering data from turbid or even opaque materials. 

Of comparable importance to these advances in experimental methods was the emergence of computer simulation as a reliable tool. The first simulations of hard-sphere liquids (effectively, colloids) were done in the late 1950s, with serious work on polymer statistics initiated in the 1960s; on polymer dynamics in the 1970s and 1980s; and on self-assembly in the 1980s and 1990s. All these fields have, alongside many others, fed on the growth in computational power available. However, even today, many important issues in soft condensed matter are too complex to be resolved by computer simulation alone.

A final reason for strong progress in soft condensed matter was the development of improved theoretical methods based on advances in statistical mechanics. Some of these were general and far-reaching, such as the scaling and renormalization-group techniques for studying phase transitions which emerged in the 1960s and 1970s (see Chapter 10). Other theoretical developments, such as the tube model of polymer entanglement (described below) were devised for specific problems in soft matter and had neither precedent nor analogue in other fields of physics. Throughout the second half of the 20th century, by a judicious combination of general and specific theoretical methodologies, an increasing number of physical properties of soft condensed matter thus came to be explained, or even quantitatively predictable, in terms of a few unifying concepts such as entropic elasticity and Brownian motion. Indeed, from the early studies of the 1940s and 1950s to the present day, soft condensed materials such as colloids and polymers have offered an important testing ground for emerging ideas in statistical mechanics. This symbiosis between soft matter experiment and statistical mechanics theory is a recurrent theme in what follows.

\section*{Condensed Matter Physics}
Condensed matter physics addresses the behaviour of systems containing very many particles, at high
enough density that each interacts with several others. Examples include
crystalline
solids and liquids. Because each atom or molecule interacts with several
neighbours, and each of
these with several more, and so on, the entire system of particles is
coupled together. This imparts
special `collective' properties to the material, such as the elasticity of
a crystal, or the
viscosity (thickness, resistance to flow) of a liquid.  Such properties
cannot usually be
understood by thinking about single particles in isolation, or even small
clusters of particles;
condensed matter physics is, in essence, `the $10^{23}$ body problem'.

During the first half of the 20th century, tremendous progress was made in
the study of condensed
phases, such as regular crystalline solids, by applying the (then)
newly-discovered laws of quantum
mechanics to periodic assemblies of atoms. This work continues today, with
ever more sophisticated
techniques being employed (see Chapters 11.1, 11.2, 11.5).
Quantum mechanics alone was not enough, however. For example, the
understanding of
magnetism in solids required not just quantum theory but also the tools of
statistical physics.
These are the rules which allow one to calculate the probability that
a many-body system is in a particular microscopic physical state (a
specific quantum state, or
`microstate') at a given temperature. The collective properties of a
condensed phase are then
expressed as averages over this probability distribution. Sometimes abrupt changes in these properties can result from small changes in control parameters such as temperature; when the abrupt behaviour becomes ever more singular as the thermodynamic limit (of large system sizes) is taken, such phenomena are called phase transitions (See Chapter 10).

Because, in thermal equilibrium, a large system explores many microstates,
its behaviour
depends on how many of these microstates there are in any given energy
range, as well as the
physical character of the microstates themselves. For example, the fact
that a ferromagnetic
material (spins on a lattice) entirely loses its magnetization when heated
above a critical
temperature (the Curie temperature) is understood qualitatively as follows.
At low temperature, the
system remains in energy states of low energy $E$, all of which are
magnetized (the spins, on
average, are aligned along a common axis). At high temperature, the system
samples states more
widely ranging in energy, the vast majority of which are not magnetized
(unaligned). The
disappearance of magnetization marks the victory of the tendency to
maximize the entropy $S$ (which
is, quantitatively, the logarithm of the number of microstates that the
system can sample) over the
tendency to minimize energy. (The magnetization vanishes smoothly at the phase transition point but its functional form changes there, so that -- for example -- its temperature derivative is discontinuous.)

This basic competition is
quantified in the celebrated formula of Hermann von Helmholtz (1821-1894) for the free energy
\begin{equation}
F = E - TS \label{one}
\end{equation}
The thermodynamic state of equilibrium at temperature $T$ in a system of
fixed contents (and
fixed volume
$V$) is the one that minimizes $F$. This represents a concise statement of
the laws of
equilibrium statistical mechanics as conceived by Ludwig Boltzmann (1844-1902) and Josiah Willard Gibbs (1839-1903). Late 20th century research, for example in disordered
magnetism, has continued to explore the conflict between energy and entropy,
revealing a battlefield
of surprising complexity (see Chapter 11.3).

Equally vital has been the issue of dynamics -- how does a many body system approach its
equilibrium state? Can it get
trapped away from equilibrium? What happens if the system is constantly
being ``driven" by
an input of energy from outside? These issues are among those discussed, in more specific contexts, below.

\section*{Soft Condensed Matter} \label{scm}
We now turn specifically to soft condensed matter.
First, what is it? There is no agreed definition, but one candidate
is as follows: a piece of soft condensed matter is a lump of material which
strongly resists
compression, but weakly resists shear. An example is a piece of crosslinked
natural rubber (latex);
this is can easily be deformed at constant volume but, perhaps
surprisingly, its
resistance to changes in volume (its bulk modulus) is as high as many a
crystalline solid.
The following materials thus qualify as soft condensed matter:
polymer gels (Jello), emulsions (mayonnaise), viscoelastic detergent
solutions (shampoo), fat
crystal networks (margarine), concentrated colloids (paint), polymer
solutions (multigrade engine
oil) and lyotropic liquid crystals (such as the slime created when a bar of
soap is left in a
pool of water). Less obvious qualifiers are
shaving foam and beer froth, which are also relatively incompressible so
long as the trapped gas is
not allowed to escape. 

Dense colloids (e.g. the thick paste made by mixing cornstarch with water) also qualify. These can develop high resistance to shear under strong stresses (a phenomenon called shear-thickening behaviour) but are nonetheless easier to shear than to compress. The case of dry powders is slightly anomalous -- such powders can fill a range of different volumes depending on how the sample has been shaken. (This is why a packet of cornflakes is always less full by the time it reaches the shop than when it left the factory.) Powders also tend to expand on shearing (dilatancy). Hence they do not fit easily into the definition chosen above but are traditionally part of the remit of soft condensed matter, in part because much of their behaviour is closely similar to that of dense colloids.

How do all these materials differ from simple liquids, such as
water? The latter is not easily compressed, and, it could be said, weakly
resists shear. But it
does so in a purely viscous manner: the force generated is proportional to
the rate of strain.
(This is called `Newtonian' behaviour after Isaac Newton (1642-1727), who first
described it.) In contrast, most of
the above materials are `viscoelastic': their response to deformation
shows a mixture of
elastic and viscous features.  For example a polymer solution, to which a
small shearing force is
applied, will first respond elastically -- it will deform with a shear
strain linear in the applied
force per unit area (the shear stress). But after a finite time (perhaps a
second or two, or much
longer in some cases),  it will start to flow like a liquid -- with a
strain rate (not strain)
proportional to the stress. Thus arise various entertaining toys such as
`silly
putty', which bounces like a rubber ball, but, if left alone, will spread
to form a thin pancake (which
is actually a puddle) on the table. The first person to consider such
behaviour theoretically was
James Clerk Maxwell (1831-1879), who invented a simple theoretical description for it (now known as the
Maxwell model).

The distinction made above is based on time-scales: water too responds
elastically at first, but so
briefly that most experiments do not detect this. The soft materials of
interest here instead show
some degree of elasticity on time scales (milliseconds to days) that are
readily observed. Note that
many soft materials, including all the examples in the list above, do
consist mainly of a simple
fluid -- such as water -- in which  `mesoscopic' objects are suspended.
Such objects have
structure on the nanometre to micron scale, and their slow dynamics
prolongs the elastic response.
Polymers (long chain molecules) or colloidal particles (spherical or
irregular lumps of hard
matter) are important examples. In both these cases, the suspended objects
are permanent in
character: unless extreme conditions are applied, they remain intact
throughout the life of the
material. Other cases, in which the suspended objects are transitory, are
considered later.

\section*{Thermal Equilibrium}
A major achievement of the late 20th century was to understand, at least at a schematic level, the equilibrium statistical mechanics of key soft materials including polymer solutions and colloidal suspensions. In many
interesting cases, the physics is
dominated by the entropic term in Eq.\ref{one}.

\subsection*{Polymers} For example, a linear polymer chain is a long, unbranched sequence of chemically identical units, connected by bonds with some flexibility. Such a chain can be viewed, at a
large enough scale,
as simply a wiggly line: there are a vast number of microstates
(corresponding to different
sequences of configurations of bonds in the chain) whose energies are not
very different from one
another. The details of the local chemistry do not matter in this limit: the polymer's properties become `universal'.
Under these conditions, what matters in Eq.\ref{one} is the
maximization of entropy; each
chain becomes a random walk. A random walk is a path, made
up of a sequence of steps,
each taken in a random direction relative to the previous one. A
picturesque example, in two
dimensions, is the trail of a drunkard staggering away from a public bar,
with no notion of
where he has been or where he is going. The path traced out by a
polymer, now in three dimensions, is similar -- with the important proviso that no two parts of the
same chain, or of any two chains, can be in exactly the same place. The detailed effect of this `excluded volume' constraint on chain statistics is subtle but the main effect is to make the chain expand to fill a larger region of space. 

Notice that any individual
polymer is constantly exploring its vast number of microstates (by Brownian
motion -- see below);
few meaningful properties can be defined  except as an average over this
exploration. This is very
unlike the atoms or molecules in a conventional `hard' material, such as
iron, where the entropic
fluctuations can be viewed as small departures (called phonons) from a
well-defined ground state (the
periodic lattice). For polymers, as is commonplace in soft condensed
matter, the fluctuations are the physics. This makes the correct use of statistical mechanics
especially challenging, and
especially important.

Many advances in understanding polymer statistical mechanics were made by
Paul J Flory (1910-1985; Chemistry Nobel Prize 1974), Pierre-Gilles de Gennes (b.1932; Physics Nobel Prize 1991) and Samuel F Edwards (b.1928; Boltzmann Medalist 1998) among others. For example, Flory devised a clever approximate treatment of the excluded volume problem, which was later cast into a field-theoretic formalism by Edwards; de Gennes exploited this insight to connect the problem with a magnetic phase transition in a certain limit (see Chapter 10). See Flory (1953), de Gennes (1979), Doi and Edwards (1986).

\subsection*{Colloids}
Another example of entropy-dominated physics is the
hard sphere colloid. Consider a suspension of micron-sized spheres, whose
interactions
are purely repulsive and of very short range. That is, centre-to-centre
spacings less than one
particle diameter are prohibited, but there is no force between spheres at
larger separations.
Then, all the microstates that are allowed at all (those with no overlaps
between spheres) have
precisely the same energy. Hence, from Eq.\ref{one}, minimizing $F$ amounts
simply to maximizing
$S$; even the temperature $T$ is immaterial.

What is the equilibrium state? This depends on the concentration of
spheres, usually expressed through their `volume fraction'
$\phi$, which is
${4\over 3} \pi R^3 N/V$, with $N$ the number of spheres in volume $V$, and
$R$ their
radius. A naive viewpoint of entropy maximization is that it corresponds to
choosing a state of
`maximum disorder'. But this viewpoint leads one astray here: for high
concentrations  obeying
$0.545
\le \phi \le 0.74$ (nothing greater than 0.74 is possible for hard
spheres), the state of highest
entropy is found to be a regular lattice or `colloidal crystal'.
Such a crystal has remarkable properties: for example its lattice planes
cause scattering of
visible light, analagous to the Bragg scattering of X-rays commonly used to probe the structure of atomic solids. This results in a iridescent, colourful
appearance, called `opalescence'. (Opalescence is named after the
appearance of gem opals,
which are also regular arrays of hard spheres, that were once in a
colloidal state
but have since dried out.)

The first suggestion that hard spheres could crystallize in the absence of attractive interactions was made in the late 1930's by John G Kirkwood (1907-1959) and the first `experimental' evidence was in fact from the computer simulation pioneered in the late 1950s by Berni J Alder (b.1925, Boltzmann Medal 2001). A closely related issue, the occurrence of orientational ordering in hard-rod fluids, was resolved by Lars Onsager (1903-1976, Chemistry Nobel Prize 1968). For more detail of these developments, see Lekkerkerker (2000).

\begin{figure}
\begin{center}
\leavevmode
  \epsfxsize=7.4cm
  \epsfbox{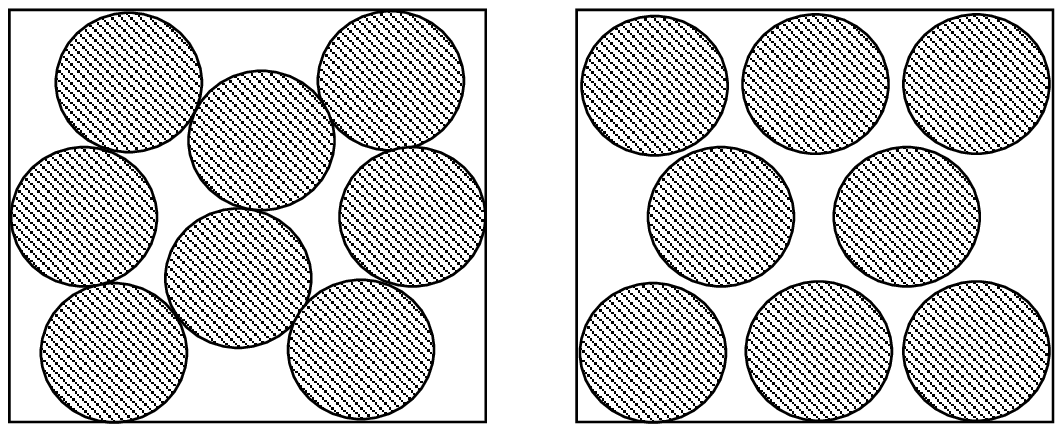}
  \hspace*{0.5cm}
  \epsfxsize=4.20cm
  \epsfbox{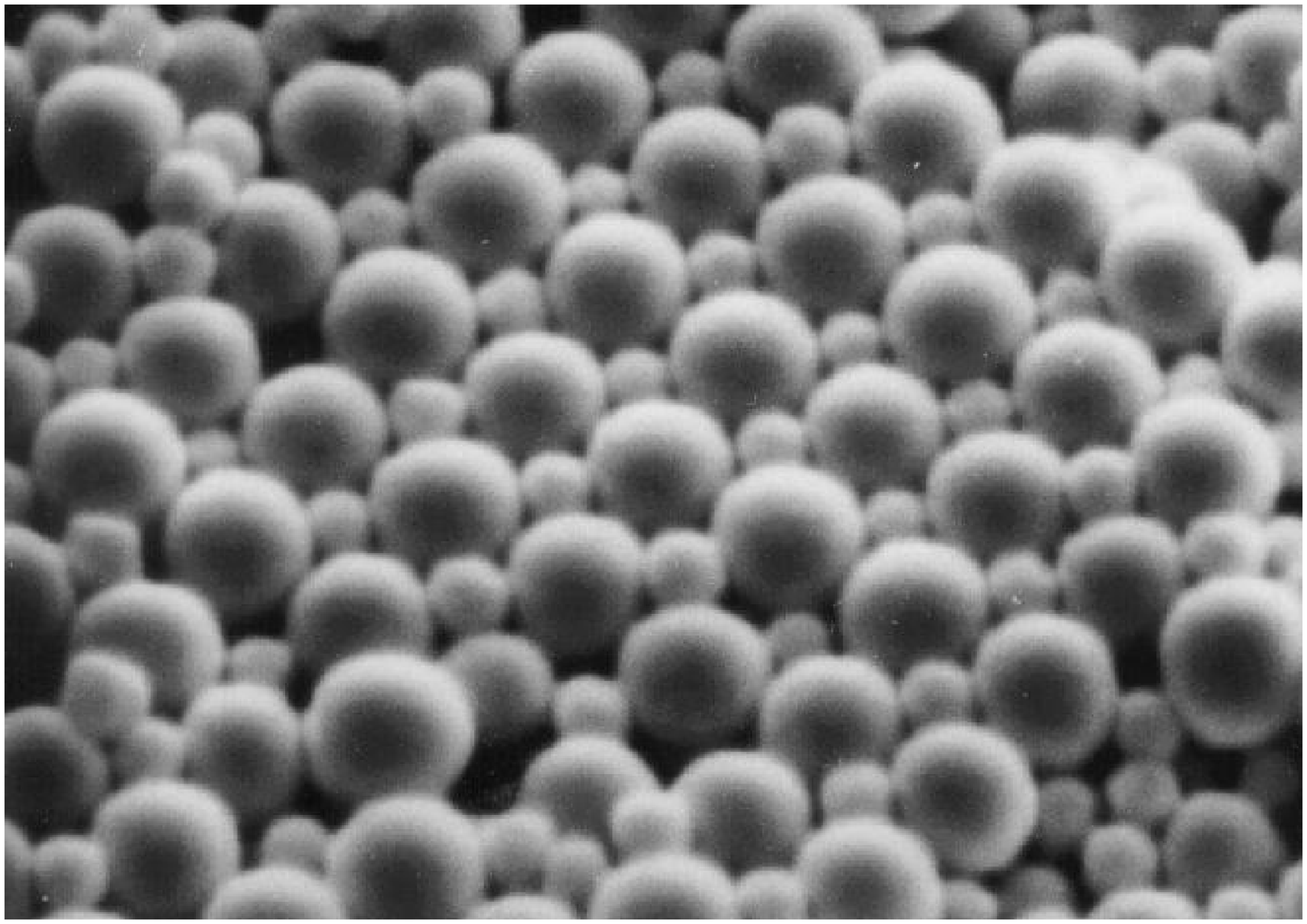}
\end{center}
\caption{In a colloidal fluid phase (left), the mean particle density is
uniform but individual
spheres have little scope for exploration without requiring neighbours to
move. In the colloidal
crystal (centre), the decrease in entropy associated with a nonuniform mean
density is compensated
by a greater local volume that each particle can independently explore.
(This is related to the fact
that a higher density is achieved in a close packed crystal than in a
random close-packed
arrangement of spheres.)  Right: a scanning electron micrograph of a dried
colloidal crystal
comprising spherical particles of poly(methylmethacrylate) of two sizes
(radii 140 and 325 nm).
Such crystals form over a few days from an initially fluid state.
(Micrograph courtesy of Andrew
Schofield, University of Edinburgh.)
\label{fig:fig5}}
\end{figure}

How can maximization of entropy lead to an ordered phase? To answer this,
we can
identify two distinct contributions to the entropy. The first one is the
entropy associated with the mean particle density $\rho({\bf r})$ which
depends on position $\bf r$.
This contribution,
$S_1 = -k \int \rho \ln \rho \, d^3{\bf r}$ is maximum for a
uniform mean density, $\rho = $ constant, as arises in the fluid state.
(Here $k$ is Boltzmann's
constant.) The crystal phase, in which
$\rho$ is strongly peaked at the lattice sites, loses heavily in terms of
$S_1$. But there is a
second entropy contribution, which is much harder to calculate: a
collective term $S_C$. This
term reflects the difficulty particles have in locally avoiding their
neighbours. In a dense
colloidal fluid, few of the spheres can be displaced a significant distance
without
also requiring many neighbours to move. Put differently, the motion of the
spheres in the fluid
state is highly correlated; this means that the number of microstates
(or distinct sphere
configurations) actually available at a given $\phi$ is much less than
$S_1$ might suggest. If, on
the other hand, the spheres occupy an ordered lattice (on the average),
then each has much more
freedom to make small excursions from its lattice site, without interfering
with its neighbours,
than in a fluid of the same volume fraction. Qualitatively, this is because
the long-range order of
a crystalline lattice is very good at packing spheres efficiently, and more
room remains for local
exploration. So, $S_C$ is much larger for the crystal than the fluid, if
the density is high. The result is that when
$\phi$ is high enough,
$S_1+S_C$ is larger for the crystal than for a fluid phase, so the crystal
is stable.
Similar arguments apply to more complex examples, such as a mixed
suspension of colloidal spheres of
different sizes (see Fig.\ref{fig:fig5}).

Although the hard sphere colloid is suspended in a solvent, for the purposes of evaluating the free energy $F$, the latter can almost be ignored. (It contributes a large but almost constant amount to the free energy.) Thus, for idealised hard spheres at a given volume fraction, the minimisation of $F$ can be performed as if the spheres were in a vacuum; this was tacitly assumed above. Such calculations were first done in a different context, however -- as an idealised model of simple atomic fluids. But in fact colloids are a much better approximation to hard spheres than atoms are; hence experimental work on colloidal suspensions has (since about 1980) proven a fertile testing ground for classical theories of the liquid state and of crystallisation. 
As discussed below, by doing experiments on suitable mixtures (e.g. colloid + polymer) one can tailor the colloidal interactions to test these theories in more detail, for example by adding an attractive component to their interaction. 

This `colloids as model atoms' approach is reviewed comprehensively by Pusey (1991); see Anderson and Lekkerkerker (2002) for a more recent account.

\subsection*{Entropic Elasticity}
The materials mentioned so far (gels, emulsions, colloids etc.) are all
soft -- but just how soft?
What causes them to have any elasticity at all? Why is the static shear modulus
not zero, as in a simple
fluid?

The first satisfactory answer to this question was given for the case of `elastomers', such as cross-linked rubbers and polymer gels. An elastomer consists of
a network of polymer
strands (each a random walk) linked together at junction points. If the gel
is suddenly deformed at
constant temperature, the random-walks strands must change their average
shape: they become
elongated along the stretching direction. The entropy of a set of deformed
strands is lower than
that of an undeformed set, for, by definition, the original
random walk is the most random state possible. (In this case the naive
argument, that maximum
randomness equates to maximum entropy, turns out to be correct.) Hence the
Helmholtz free energy
$F$ increases on applying a strain. The amount it increases is calculable in
statistical models; in a small shear strain, characterized by a small
strain angle
$\gamma$ (measured in radians) the free energy change is $\Delta F \simeq
kTN\gamma^2$, where
$k$ is Boltzmann's constant, and
${N}$ is the number of network strands. Equating this stored free
energy to $V G\gamma^2/2$ defines the shear modulus $G$; this is of order
$kT{N}/
V$, or, $kT$ times the number of `elastic degrees of freedom' ($N$) per
unit volume. Values for
a typical polymer gel lie in the range $G = 10-10^4$ Pa, depending on the
concentration of
polymer and the amount of crosslinking. The upper end of this range lies
several orders of magnitude
below the shear modulus of, say, a typical metal. And, at the lower end, a
value
$G\simeq 10$ Pa represents a material whose elasticity resembles that of
half-cooked egg-white.

These order-of-magnitude estimates for the modulus of elastomers stem from the pioneering work of Flory and others (Flory 1953). From the 1970's onward, sophisticated quantitative theories for their elastic behaviour were developed to allow e.g., for the effect of `entanglements'. The latter arise from the fact that polymer chains cannot pass through each other; the result is an effective shortening of the network strands and an increase in their number.

Still smaller values ($G \le 1$ Pa) can arise for the case of colloidal
crystals. For
these, a similar formula applies:
$G \simeq kT{N}/V$ where ${N}$ is the number of spheres.  Compared with
such a tiny modulus,
even the force of gravity, acting on a sample a few millimetres across, is
strong. A sample larger
than this will usually collapse under its own weight, giving a nearly-flat
meniscus in a test-tube,
so that to the untrained eye it will appear to be a fluid (Pusey 1991).

In both elastomers and colloidal
crystals, the smallness of the shear modulus $G$ stems from the small number
of elastic degrees of
freedom (chains or spheres) per unit volume. In a conventional solid the
elastic degrees of freedom
correspond to atoms or small molecules, and the number of these objects per
unit
volume is vastly higher, giving $G$ values of order $10^8$ Pa or more.
Note that similar arguments do
not apply to the bulk modulus
$K$, which controls the free energy response to volume changes $\Delta V$
($\Delta F = {1\over 2} K
(\Delta V)^2$) in a system at fixed contents. The bulk modulus is
dominated not by the polymers or colloids, but by the surrounding liquid,
which is water, or some
equally `normal' molecular material (having, typically, $K \ge 10^8$ Pa).
On the other hand, the experimenter can choose to study, instead of $K$, the
`osmotic modulus'
$\bar K$. This is defined (and measured) by confining a soft material
within a semipermeable cell of
variable volume. In this case the system's contents are not fixed: solvent
can enter and leave,
though the polymers or colloids cannot. The solvent contribution to the
bulk modulus is
thereby eliminated, and one finds that
$\bar K$ is again very small, roughly comparable to $G$.

\subsection*{Mixtures} \label{mixt}
In many real-world materials, both polymers and colloids are simultaneously
present. Indeed, they
may be accompanied by emulsion droplets, or surfactant aggregates
(discussed further below), or yet further types of suspended object, in
almost any combination.
These components can interact in a strongly non-additive way: such a
material is much more than the
sum of its parts.

\begin{figure}
\centfig{12cm}{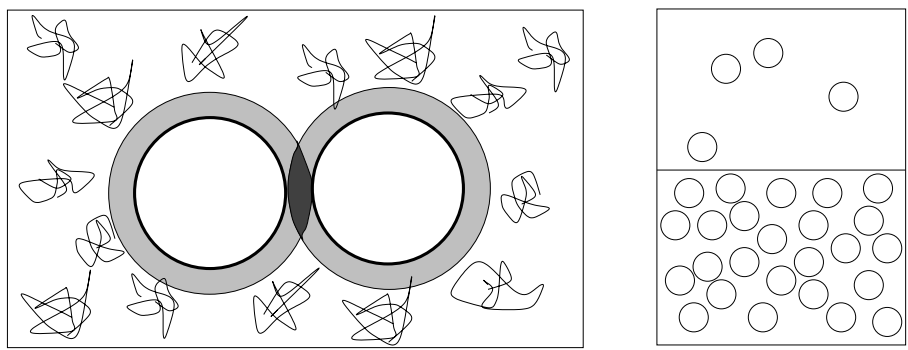}
\caption{The ``depletion force" in a colloid/polymer mixture. The polymers
avoid the exclusion zone
around each sphere (light shaded). By overlapping these areas, the total
volume still available to
the polymers can be increased by an amount equal to the dark shaded region.
The resulting increase
in polymer entropy causes a reduction in free energy of states in which the
colloidal spheres are
close together, that is, an effective attraction
between spheres. This is called the depletion force. Like the attractive
(van der Waals) force
between intert gas atoms, it can cause a gas/liquid phase separation: in
this context,
coexistence of colloidal fluids of different volume fractions (right). The
polymers (not shown)
have a higher concentration in the phase with less colloid.
\label{fig:depletion}}
\end{figure}

For example, suppose large colloidal spheres are mixed, in solution, with
small polymers. The
polymer coils cannot approach the impenetrable colloids too closely without
becoming flattened out
along the surface -- a deformation which would cause them to lose entropy.
To avoid this, the
polymers tends to stay out of an `exclusion zone' around each colloid;
this also loses the
polymers some entropy, though not as much.
But now an even more economical arrangement is possible:
this is to overlap the exclusion zones of different colloids.  Such
overlaps mean that
the total `excluded volume' (the volume that the polymers cannot get into
without
strongly deforming) is reduced; so the polymer entropy increases (Fig.
\ref{fig:depletion}).
This argument shows that the free energy of the polymers is minimized by
states in which the large
colloidal spheres are close together. The effect of this is just as if
there were an attractive
force between the  spheres. The range of this polymer-mediated `entropic
attraction' depends on the
mean chain size; its strength, on the polymer concentration.

As mentioned previously, the physics of the colloidal particles
in such
a mixture is  closely analagous to a system of inert atoms (such as argon).
These have a
strong repulsion when their electron clouds overlap, but a weak attraction
at slightly larger
distances. Indeed, by tuning the polymer concentration, analogues of the
various phases of an
atomic system can be mapped out experimentally. As well as the transition
between a colloidal fluid
and colloidal crystal (already described above), a `gas-liquid' transition
is now found, between
two colloidal fluids having different volume fractions $\phi$. By carefully
varying the polymer
size, studies have shown that the liquid state only exists when the
attractive interactions
are of long enough range. Such work is important for understanding
colloids. It also illuminates a
fundamental issue of molecular physics: the origin of the liquid state itself (Pusey 1991, Anderson and Lekkerkerker 2002).

\subsection*{Coulomb and Dispersion Forces} \label{coulomb}
In many soft materials
Coulombic, {\em i.e.}, electrostatic, interactions are vital:
polymers, colloids and surfactants can all contain chemical groups which
ionize when placed in
water. Coulomb forces rank alongside entropy as a major determinant of
physical properties, at
least in aqueous systems, and especially in biological ones. The same applies to dispersion forces -- attractive interactions caused by the correlation of fluctuating dipole moments on different atoms or molecules.

Theories of simple ionic solutions (such as sodium chloride) were developed in the early 20th century. These were later combined with dispersion forces in theories of surface interactions by Lev D. Landau (1908-1968; Physics Nobel Prize 1962) and others; the resulting theory of colloidal interactions and its detailed experimental test are reviewed by Israelachvili (1985). 
By the end of the 20th century, the outstanding problems in this area mainly concerned the effects of strong correlations between charges; for example, current theories cannot adequately cope with
`macroionic' systems in which a polymer chain, or a colloid, carries hundreds of charges whose positions are thereby strongly correlated. In some regimes of strong charge correlation, even the basic sign of the interaction (attractive or repulsive) remains unresolved at the time of writing.

\section*{Dynamics}
Even when the suspended objects are permanent (as for the
colloidal and polymeric systems so far described), and all the same size,
their dynamical behaviour
can be somewhat complicated. In particular,
the response of the material to a flow depends on how quickly its internal
structure can reorganize
to adapt to the changing shape of the sample.

The basic
mechanism of this reorganization is Brownian motion: the jostling of a
suspended object by
continuous random collisions with the molecules of the surrounding fluid.
Brownian motion is named
after the botanist Robert Brown (1773-1858) who first observed it in pollen grains; it was
explained by
Albert Einstein (1879-1955; Physics Nobel Prize 1921) in 1905.
(The effect was used by Jean Baptiste Perrin (1870-1942) to determine
Boltzmann's constant $k$; for this he received the 1926 Physics Nobel Prize.)
Brownian motion in soft condensed matter is complicated by the facts that (i) the suspended objects
collide  
with each other as well as with the solvent; and (ii) for flexible objects
such as polymers,
collisions cause changes in shape, as well as in position, of these
objects.

\begin{figure}
\centfig{12cm}{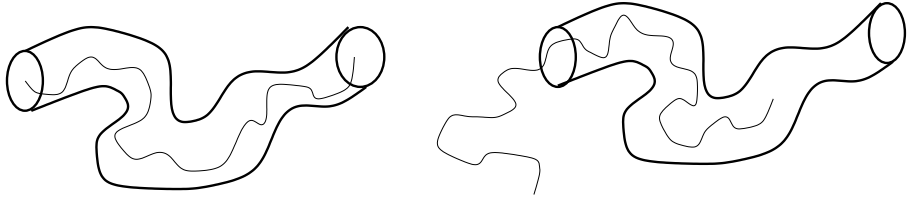}
\caption{The tube model. In an entangled polymer system, a given chain
(light curve) is densely
enmeshed with its neighbours. The effect of the collisions with these
neighbours is to
confine the Brownian motion of the chain to the tube-like region depicted
here. However, the chain
can still disentangle itself by creeping along the tube axis (right). As it
does so, it becomes
enmeshed in a new tube (not shown) whose orientation is uncorrelated with
the first. The process,
called curvilinear diffusion, can be fully quantified within the model, and
from this the stress
memory function $\mu(t)$ can be found. (In fact, $\mu(t)$ is simply the
fraction of the initial tube
from  which the chain has not escaped by time $t$.) In strong flows, the
tube becomes deformed and
this enhances the rate of disentanglement. These effects are quantified in
the Doi-Edwards
constitutive equation relating the state of stress $\sigma_{\alpha\beta}$
to the deformation history
of the medium. A simplified version of this (called the `independent
alignment approximation' or
IAA) reads
$\sigma_{\alpha\beta}(t) = G\int_{-\infty}^t (d\mu(t-t')/ dt')
Q_{\alpha\beta}(E_{\gamma\delta}(t,t')) dt'$ where
$Q_{\alpha\beta}$ is a specified tensorial function of the cumulative
deformation tensor
$E_{\gamma\delta}(t,t')$ between an earlier time $t'$ and the present time
$t$ (whose logarithmic
derivative is the strain rate tensor $\kappa_{\gamma\delta}$). The IAA
equation is less accurate
than the full equations of Doi and Edwards (1986), which are too complicated to reproduce here.
\label{fig:tube}}
\end{figure}

For example, polymers in solution, above a certain concentration, become
hopelessly
entangled with one another: the system resembles overcooked spaghetti (or,
indeed, a
`can of worms'). The flow behaviour of a polymeric fluid is  dominated by
the slowness with which
these entanglements can be unravelled by Brownian motion. A mechanistic
proposal for this, first put forward de Gennes, was developed by Edwards and Masao Doi (b.1948) into
a simplified but quantitative theory of polymer viscoelasticity, called
`the tube model' (Doi and Edwards, 1986). Each
polymer chain is envisaged as confined to a tube formed by entanglements
with its neighbours;
collisions keep it within the tube, but Brownian motion still allows it to
creep slowly along the
tube axis (Fig.\ref{fig:tube}).

Within the
tube model, one can explain what happens in an experiment like the following one. Take a soft polymeric
material; suddenly shear it through a small angle
$\gamma$ which is then held constant; and measure the resulting shear
stress, $\sigma(t)$, as a
function of the time $t$ since the strain was applied. For an elastic
solid, such as the crosslinked
polymer gel considered above, the stress forever obeys
$\sigma = G\gamma$. For a viscoelastic material, one finds $\sigma = G \gamma \mu(t)$
where
$\mu(t)$ is a decaying function of $t$. By convention $\mu(0)
= 1$ so that $G$ represents an `instantaneous' elastic modulus, and
$\mu(t)$ its decay. The tube
model predicts, to reasonable accuracy, the shape of $\mu(t)$ and its
dependence both on the mean
chain length and on the state of entanglement.

Much more significantly, it also predicts the
nonlinear viscoelastic behaviour of the system at {large} strains. In
polymers this is characterized
by a `shear thinning' effect: if a large strain is applied (shearing through an angle $\gamma$ that exceeds, say, 0.1 radian)
then the material becomes, temporarily, even softer than before. More
generally, the predictions of
the tube model are summarized in a `constitutive equation' which relates
the state of stress at
time
$t$ (formally the stress tensor
$\sigma_{\alpha\beta}(t)$) to the history of deformations previously
applied (formally the velocity
gradient tensor
$\kappa_{\alpha\beta}(t'<t)$). The Doi-Edwards constitutive equation has
been very successful in
predicting many phenomena previously unexplained. These include, for
example, the observation that
if a polymer solution is stirred with a rod, its top surface will climb up
the rod. This is the
exact opposite of what a Newtonian fluid does under the same conditions.

The tube model (and subsequent, more refined models based directly upon it)
sets an enduring standard for what a good dynamical theory of soft
condensed matter should aim to achieve. It offers a simplified but quantitative understanding of the relation between material properties and the underlying 
molecular disposition. To date, however, there is no other class of soft materials for which comparable success has been reported. One reason for this is that the tube model exploits the fact that, even under fairly strong flows, entangled polymer chains remain close to equilibrium on the mesoscopic scale that defines the tube. In many other soft materials, such as dense emulsions (mayonnaise) the same is not true: the strong flow regime brings strong departures from equilibrium even at a local scale. The experimental and theoretical situation is comprehensively described in Larson (1999).

\subsection*{Self-Assembly} \label{sa}
Let us now consider the case when the mesoscopic objects in a fluid are not
permanent, but transitory in character, such as can arise by the `self-assembly'
of small molecules into large clusters.

\begin{figure}
\centfig{11cm}{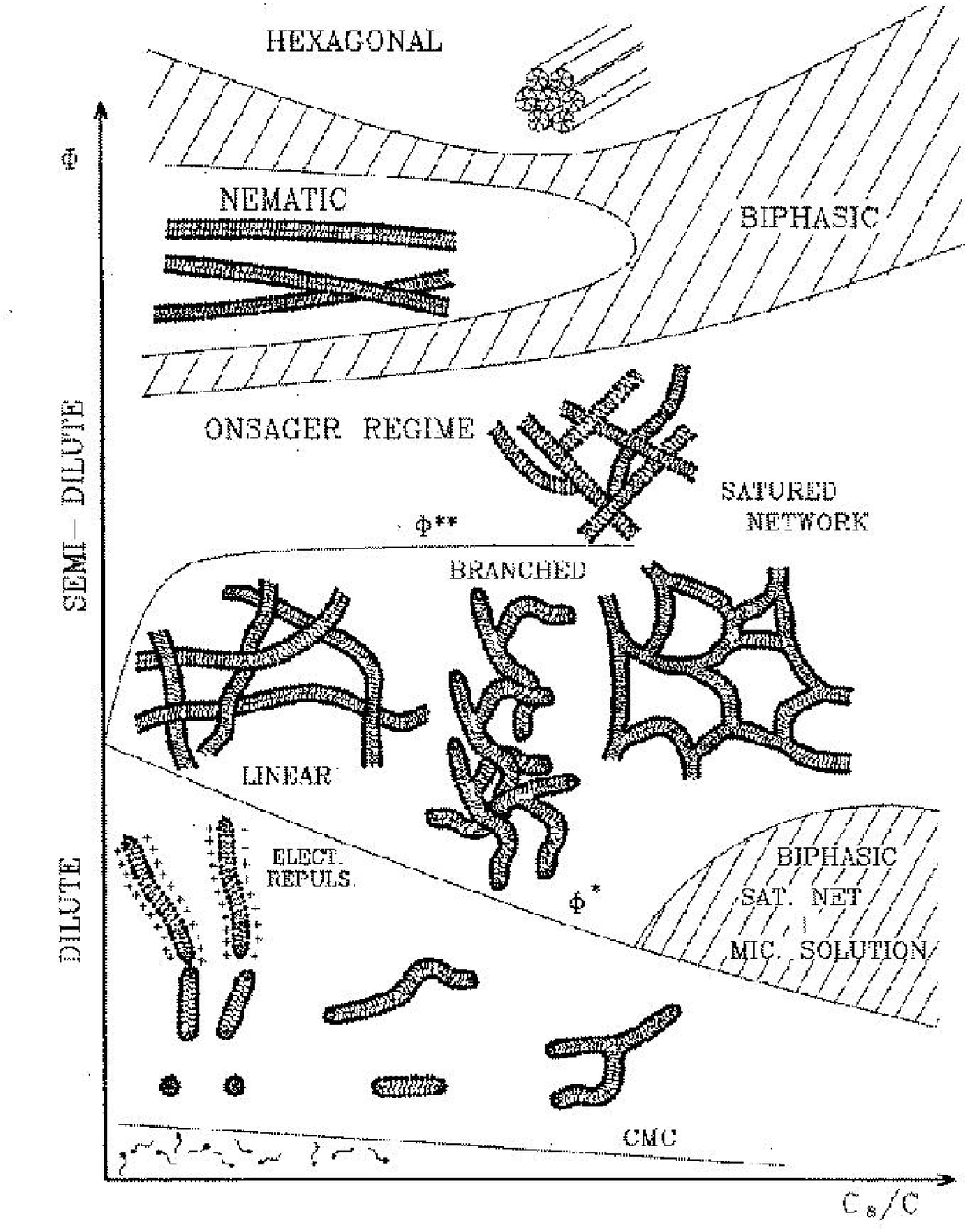}
\caption{Schematic phase diagram of cylindrical aggregates in water.
Regimes of different
organization are sketched on the $C_s/C,\Phi$ plane, where $C_s/C$ is the
molar ratio of salt to
amphiphile, and $\Phi$ is the volume fraction of amphiphile. Shaded areas
are where the equilibrium
state of the system is represented by coexistence of different phases
(compare the colloidal
vapour/liquid coexistence shown in Fig.\ref{fig:depletion}). Lines marked
$\Phi^*$ and $\Phi^{**}$
denote gradual crossovers between a dilute regime, the so-called
`semi-dilute' regime of entangled
(and/or branched) flexible cylinders, and an `Onsager' regime of entangled
semi-stiff cylinders,
respectively. At still higher concentrations, two liquid crystal phases
(nematic and hexagonal)
arise.  For ionic detergents, electrostatic interactions are important at low
$C_s/C$ and low
$\Phi$, as indicated. Below the line marked CMC (at very low $\Phi$ values)
the detergent is present as a molecular solution; there is no longer
significant aggregation. (Figure courtesy of Fran\c cois Lequeux and Jean
Candau.)}\label{fig:cylind}
\end{figure}

For example, a simple detergent
molecule (or soap) has a hydrocarbon tail, which repels water, capped with
a water-loving head
group. This type of molecule is called an `amphiphile'; their propensity for self-assembly and for adsorption at surfaces fascinated Irving Langmuir (1881-1957, Chemistry Nobel Prize 1932).
When placed in water, such molecules will spontaneously form clusters, or
`aggregates', that
minimize the contact area between the tails and the surrounding liquid.
These aggregates are
transitory, since Brownian motion is enough to break them. (In many cases
they flicker in and out
of existence on a microsecond timescale.)  Depending on the relative sizes
of head and tail,
the optimal packing can be a small sphere (called a micelle), a long
flexible cylinder (giant
micelle), or a flat sheet (bilayer). Giant micelles are objects similar to
polymers, but their
transitory nature leads to more complex dynamics (see Cates and Candau 1990): reversible splitting and
re-forming of the chains
modifies the process of Brownian disentanglement. And in some systems, the
micelles can become
branched, forming a fluid network of one-dimensional, cylindrical tubules.
These and other states
of organization for cylindrical aggregates are shown in the schematic
`phase diagram' of
Fig.\ref{fig:cylind}.

\begin{figure}
[htbp]
\begin{center}
  \leavevmode
  \epsfxsize=4cm
  \epsfbox{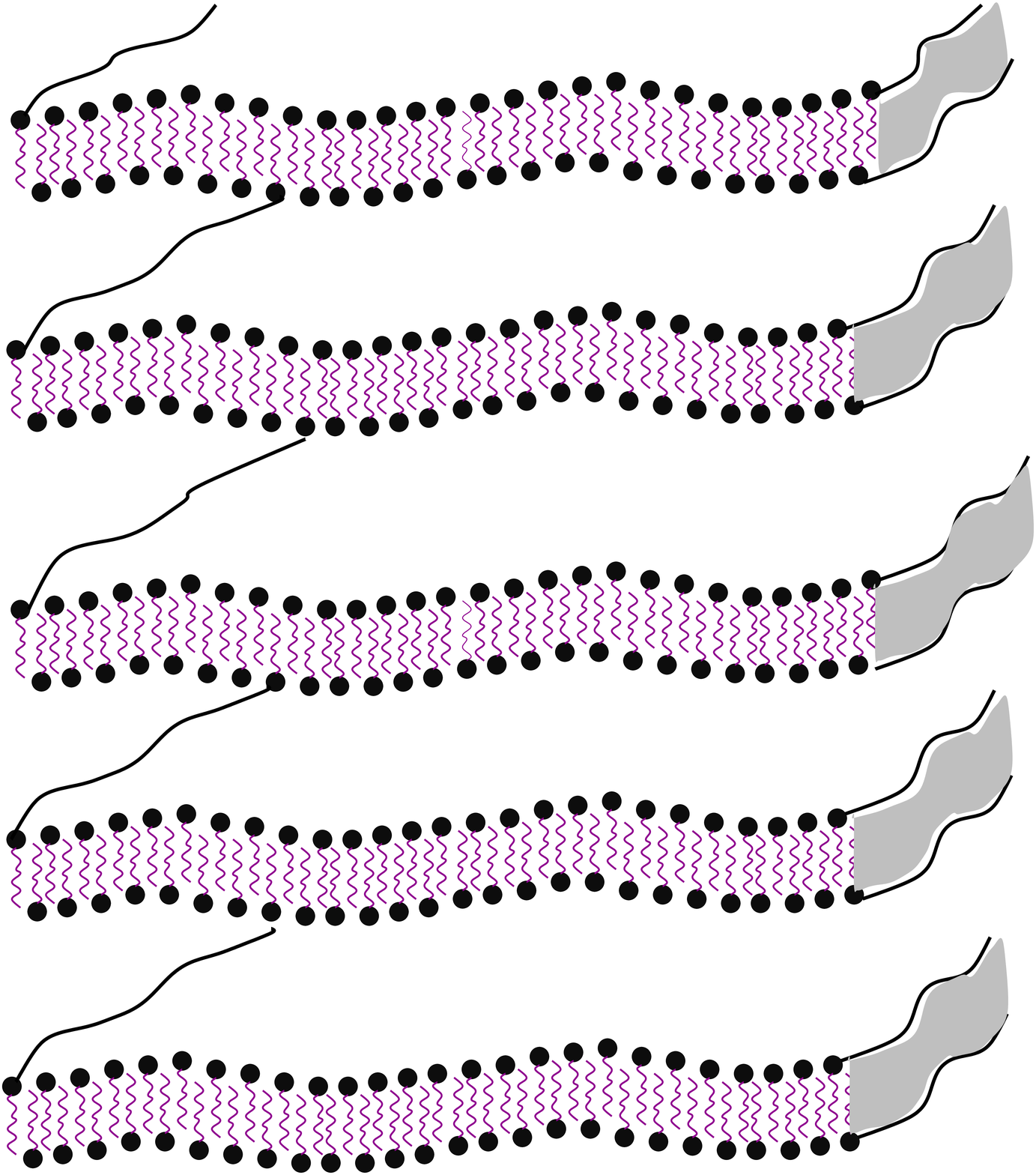}
  \hspace*{1.5cm}
  \epsfxsize=4cm
  \epsfbox{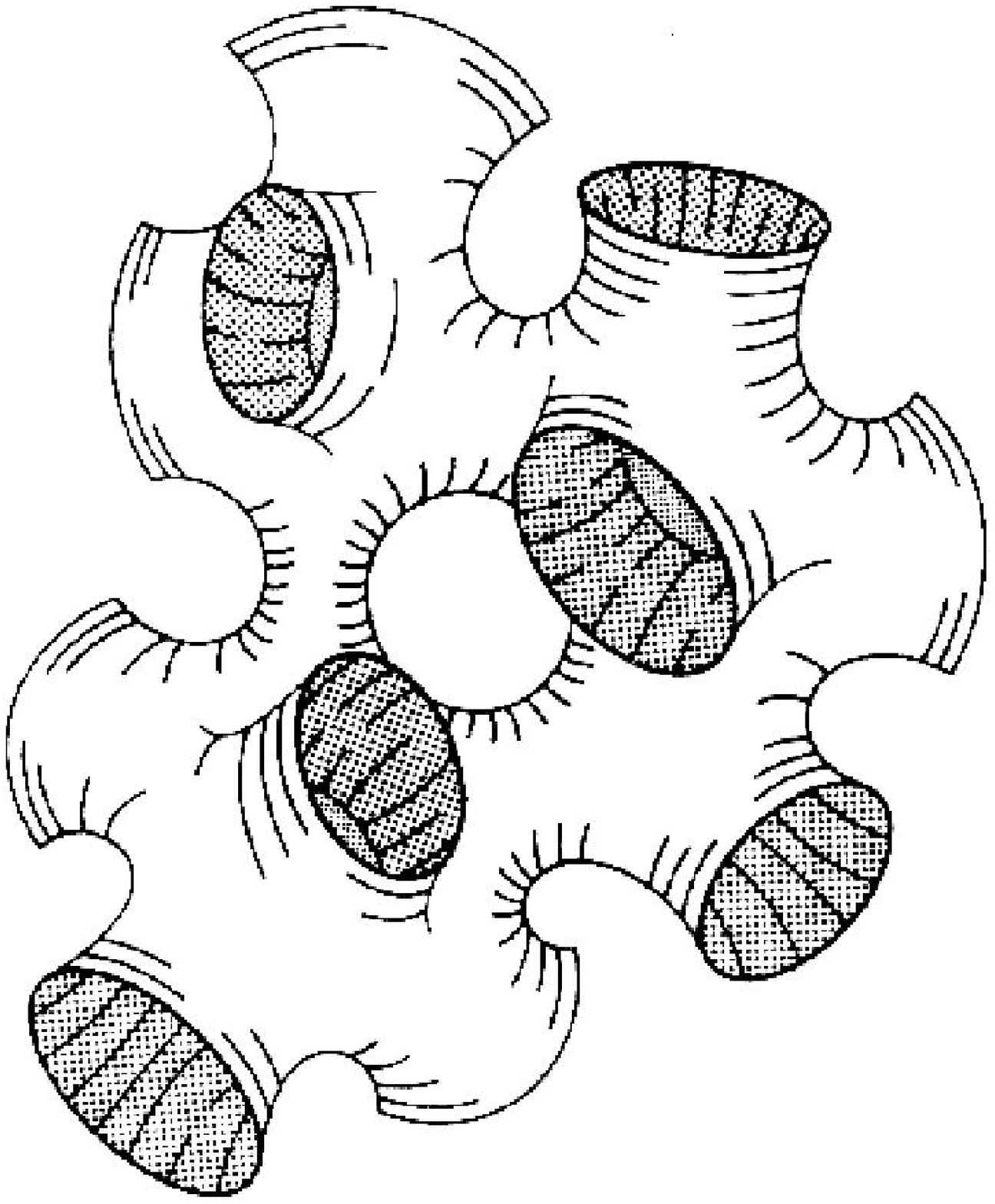}

\end{center}
\caption{States of organization for bilayer films. Both outer surfaces of
such a film
are coated with  head-groups of the amphiphilic molecules; these are
exposed to water and protect
the inner regions where the tails reside. In the lamellar phase, extended,
roughly flat, bilayer
sheets form a stack (left); in the sponge phase (right) these fuse into a
single surface of complex
topology. (Drawing: courtesy Gregoire Porte.) Such a bilayer surface
divides space into two distinct
solvent domains, which may or may not have equal volumes (symmetric or
asymmetric sponge).
\label{fig:bilayer}}
\end{figure}

Turning now to the
case of an extended bilayer (Fig.\ref{fig:bilayer}) this can reconnect
itself topologically to form
a multiply-connected, sponge-like film on which the detergent molecules
live as a two dimensional
fluid. There is a surprising, and deep, analogy between this film and the
interface between `up'
and `down' spins in a ferromagnetic solid. The bilayer film divides space
into two domains which
are of equal volume at high detergent concentrations (the `symmetric
sponge', analogous to a
paramagnet) but of unequal volume at lower concentrations (the `asymmetric
sponge', analagous to a
ferromagnet).  In some bilayer systems these two distinct phases, both of
which are isotropic
fluids, are connected by a phase transition where various
anomalies, for example in
turbidity and light scattering, have been observed.

The same local structures of cylindrical or sheet-like aggregates can
also arise in ordered phases, such as a regular array of cylinders, or a
regular stack of sheets;
both are liquid crystals. The cylinders show periodic order (crystallinity)
in two
directions but fluid-like behaviour in the third, which lies along the
cylinder axis. In the
bilayer case, extended two-dimensional fluid films adopt a pack-of-cards
arrangement, with
periodic order in the third (stacking) direction. This  so-called
`lamellar' phase is actually
the structure of wetted soap slime, mentioned previously; its very
sliminess stems from the ease
with which the layers can slide over one another (just like a deck of cards).

In the case of cylindrical micelles, a third
possibility exists, called the `nematic' liquid crystal. In this phase,
the micelles all point
in roughly the same direction, spontaneously breaking the rotational
symmetry of an isotropic fluid,
but without long-range positional order in any direction.

All these liquid crystal phases have special kinds
of elasticity which reflect their unusual state of order; their properties have been explored by Landau, de Gennes and Onsager among others (see de Gennes 1972). And, just as in
the polymeric and
colloidal systems described above, entropy and thermal fluctuations play an
important role. For
example, in the lamellar liquid crystal, Brownian motion causes each fluid
bilayer to
wobble slightly around its mean position. These fluctuations are restricted
by the fact that each
bilayer cannot cross its neighbours. The more flexible the bilayers, and the closer they lie
together, the larger is this entropic penalty. So, while a stack of very
flexible bilayers is easily
bent, it is much harder to compress (even if the solvent is allowed to
escape -- an `osmotic compressibility' test). This is,
incidentally, why the onion-like bilayer structures shown in Fig.
\ref{fig:onions} (and discussed
below) are polyhedral, not spherical, in shape:
although a sphere minimizes
curvature, a polyhedron, by making the best use of space, minimizes
compression. 

The osmotic compressibility and other properties of self-assembled phases, can in many cases, be quantiatively related to the local elastic constants of the aggregates that they contain. These elastic constants are few in number and their control of macroscopic properties represents a striking example of the universality concept mentioned in the introduction. This key insight was the basis of pioneering work by Wolfgang Helfrich (b.1932) who applied these ideas in the 1970's first to lipid membranes arising in a biophysical context; the approach is reviewed, alongside many of the subsequent developments outlined above, by Gompper and Schick (1994). Since the early 1990s, the study of self-assembled amphiphilic systems increasingly embraced biophysical applications, with simplified lipid bilayers extensively studied as models for cell membranes; such work is reviewed in Lipowsky and Sackmann (1995).
 
\subsection*{Fractal Aggregates and Colloidal Gels}
A somewhat different type of spontaneous aggregation occurs in colloidal
suspensions when attractive forces are present between colloidal particles
(sticky spheres). The particles can adhere to form tenuous threadlike structures, typically `fractal' in nature. A well-known example is the domain structure in a ferromagnet at the critical point. Here regions of positively and negatively aligned spins form fractal patterns, so that, as long as one does not look closely enough to resolve individual spins, it is impossible to tell from a picture of the spin pattern what magnification or resolution has been used. Similar remarks apply to photographs of colloidal aggregates formed, for example, by suddenly changing temperature or pH to ensure that a strong attraction (e.g. from dispersion forces) arises between colloidal particles that were previously repulsive to one another (e.g. because of a Coulomb repulsion). 

The realisation that fractal concepts could be used to describe colloidal aggregation caused an outburst of interest in the problem among physicists in the 1980s. This followed from pioneering work of Thomas A Witten (b.1944), who first recognised that the nonequilibrium fractal structures seen in simple models of aggregation were signatures of a critical phenomenon analagous to that found at continuous phase transitions (see Chapter 10). It is notable that computer simulation played a key role in this work; direct comparison between experimental and simulated data became the main test of theory in this area. These developments are reviewed by Vicsek (1989). 

Whenever these fractal aggregates
link up into a continuous web, a network arises, called a `colloidal gel'. 
Depending on the strength
of the interactions, the
aggregates may or may not gradually break up and reconnect by Brownian
motion; often, they
can easily be broken apart by an applied stress. This can lead to a
different type of
viscoelasticity from that described above for polymers. Here the material
remains an elastic solid
almost indefinitely ($\mu(t) = 1$) so long as the applied stress
$\sigma = G\gamma$, remains below a certain threshold. Only if this `yield
stress' is exceeded,
does the material begin to flow. To add to the complication, the yield stress itself can slowly evolve with time in a process called `ageing'. 

At the time of writing, the formation, flow and aging behaviour of colloidal gels remains relatively poorly understood -- partly because, for the gel to flow at all, the bonds cannot form in a completely irreversible manner as simple aggregation models assume; see e.g., Haw and Poon (1997).

\subsection*{Flow-Induced Phase Transitions} \label{fits}
Both amphiphilic and colloidal aggregates can show a strong coupling between
their state of organization and the state of flow in the material. This can
have spectacular
consequences.  For
example, in some cases it is possible to convert an isotropic fluid,
containing a
spongelike amphiphilic bilayer, into a lamellar liquid crystal, merely by a
light
shaking of the sample in a test tube. Similarly, many colloidal crystals
can be rapidly converted to
a fluid state by slight shearing. These are both `flow-induced phase
transitions'. (The term `phase transition' means an abrupt change in the response of a system, or the functional form of this response, under a smooth change in parameters such as temperature or -- here -- flow rate: see Chapter 10.)

In the simplest cases, the shear induces a transformation which could
also have been made by changing a thermodynamic variable such as
temperature or pressure. Indeed, in
both of the above examples, the same effect could be achieved by changing
the concentration
slightly, rather than applying the shear. Several flow-induced transitions
are now well
understood in these terms.

In a second class of systems (including giant micelles) an abrupt
transition from one state of flow to another, as a function of flow rate,
is required by
the constitutive equation. This `intrinsic' flow instability may or may
not be accompanied by
a change of thermodynamic state, for example to a hexagonal or a nematic
liquid crystal.

\begin{figure}
\centfig{12cm}{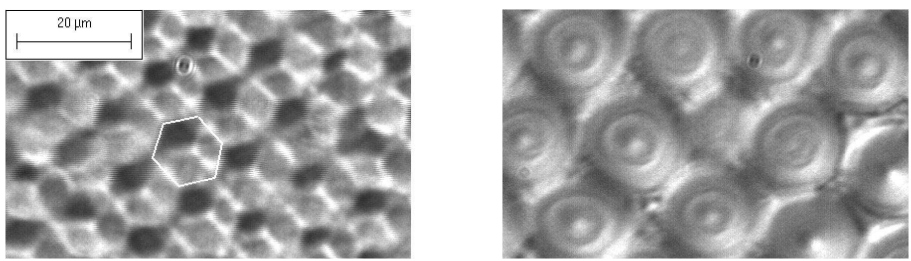}
\caption{An onion texture in a sheared lamellar phase, viewed with
differential interference
contrast microscopy. Immediately after prolonged slow shearing, the
close-packed onions have
polyhedral shapes (left). It is hard from such a picture to see where one
onion ends and the next
begins, but an example is picked out in white. If some extra solvent
(water) is now added, the
onions swell significantly in size and become more separated. Their
spherical topology is now
more clearly visible. (Figure courtesy of Mark Buchanan.)}\label{fig:onions}
\end{figure}

In a third class of systems, the application of a flow produces a state of
organization which
is entirely new. A spectacular example is when a
lamellar liquid crystal is subjected to prolonged steady shear
(Fig.\ref{fig:onions}). Rather than
remain as a regular stack, the layers slowly reorganize into
closed shells of spherical topology (but polyhedral shape) nested one
inside another, forming an
array of `onions'. (Each onion is typically a few microns across, and
contains hundreds of layers.)
This transformation is reversible: by reducing the shear-rate, the previous
structure is recovered.
However, if shearing suddenly ceases altogether (so there is no longer a
steady input of energy
from outside) the onions cannot revert to their previous structure at all
quickly; in some cases,
they are stable for months. This provides an interesting technology for
encapsulating small molecules, such
as pharmaceuticals, which can be added to the mix before the onions are
made.

Another example of a flow-induced phase transition occurs in solutions of
giant micelles, at
concentrations below the onset of entanglements. Such a solution is
barely viscoelastic, but when sheared slowly and for a long period, it
transforms into a new,
gelatinous state.  Theoretically, one expects shear to have a strong effect
in a system like this,
but only when the shear rate $\dot\gamma\tau \ge 1$, where $\tau$ is the
longest relaxation
time observable in the system at rest. (Here, $\tau$ is the rotational
relaxation time of
the micelles: micro- or milliseconds.) But in this case the
shear rate required
is tiny: $\dot\gamma\tau
\le 10^{-3}$; this is
far from understood.

The first systematic experimental studies of flow-induced phase transitions in self-assembled materials were carried out in the 1980s (Rehage and Hoffmann 1991), whereas the onion transition was not discovered until later (Diat, Roux and Nallet 1992). Despite much theoretical effort, there had been relatively little progress in understanding these transitions except for a few limiting cases, some of which are reviewed by Larson (1999).

\subsection*{Jamming and Granular Media}
Under some conditions, a suspension of hard-sphere colloids can undergo a
flow-induced phase transition from a relatively free-flowing phase into a
`jammed state'.
This appears to be
related to the physics of traffic jams: on a crowded freeway, there is a
mean speed above
which traffic cannot flow freely. Even if all drivers wish to go faster
than this, they cannot,
because the flow is unstable, and small fluctuations lead to spontaneous jams.

\begin{figure}
\centfig{8cm}{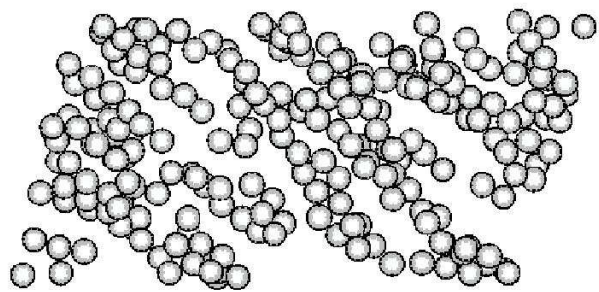}
\caption{The jamming transition in a sheared colloid. The data are from a
computer simulation of a
hard sphere colloidal suspension at $\phi =0.54$ which has been strained to
$\gamma = 0.22$. Shown
in the figure are only those spheres which have come into very close
contact ($\le 10^{-5}$
radius) with at least one neighbour. As can be seen from the figure, the
contact geometry is
strongly anisotropic and suggests the formation of `force chains' running
top left to bottom
right. Similar force chains have been reported in granular materials. (The
simulation is by J
Melrose, Cavendish Laboratory; the figure is courtesy of him.)
\label{fig:sim}}
\end{figure}

A classical experiment along these
lines can be performed in the kitchen: place a tablespoonful of starch
granules (corn starch or
custard powder) in a cup, and add water drop by drop, until the material is
only just properly wet.
Now stir with a spoon. If the composition is just right, the
suspension will
flow almost without resistance, so long as the spoon is moved slowly. But
an attempt at rapid
stirring is completely frustrated -- the whole thing sets into a solid lump.
Work in the late 1990s began to illuminate such processes (Fig.\ref{fig:sim}), although the physics of the situation is far from clear.

The nature of the jamming transition in colloids also suggests an intriguing connection with another
class of easily-jammed materials: granular media. These include dry
powders, such as sand. There is a continuous spectrum connecting dry powders (via slurries and pastes) all the way to the colloidal state. Across this
spectrum, some common tools began to emerge during the 1990s, for example, a characterization of local geometry
based on `force chains' of
strong local contacts. But a truly unified theoretical framework remains lacking.

\subsection*{Arrested Dynamics} \label{meta}
Many soft materials, such as the deformed fluid droplets in a dense
emulsion, or the onions in
Fig. \ref{fig:onions}, are trapped in an arrested
state, far from true equilibrium, and yet close enough to a metastable
minimum of free energy to
allow thermodynamic ideas to be employed locally, or for short time scales.
At longer time scales, such materials can exhibit very slow relaxation
processes; in
several cases, no matter how long one observes the system, its properties
continue to evolve.

From the 1970s onward, broadly analagous behaviour to this had been seen in other areas of condensed matter, particularly in disordered magnets called `spin glasses' (see Chapter 11.3). This work was itself partly inspired by earlier experimental studies of `structural glasses' that arise by supercooling of molecular fluids, such as glycerol, into the vitreous regime. In the 1980s, dense colloidal suspensions were themselves shown to be structural glasses, in the sense that the particles become trapped close to fixed positions although these do not form an ordered lattice (Pusey 1991). In the 1990s it was proposed that that glass paradigm could explain a wider range of arrested dynamics in soft condensed matter, such as that of dense emulsions or textured liquid crystalline phases (e.g. onions) in which one can expect a slow evolution of the system through a sequence of metastable, near-equilibrium states. It was also proposed that the jamming transition might be related to a transition from fluid to glass caused by the imposition of a stress. These and other ideas are explored in a recent proceedings (Cates and Evans 2000).

\subsection*{Phase Transition Kinetics}
A simpler form of metastability than that found in glasses arises in a supercooled
vapour, for example. In such cases the system
resides in one of a small
number of free energy minima, corresponding to different thermodynamic
phases, even though
its free energy $F$ could be lowered by moving to another one.
The transition may finally proceed, for example, via nucleation
of a small droplet of the preferred phase. A different mechanism arises
when, because of a sudden change of parameters (such as a temperature
change or `quench') a
system which was originally at a free energy minimum suddenly finds itself
at or near a maximum,
and starts to evolve accordingly. An example is when a stable mixture of
two fluids is cooled, and
starts to demix by a process called `spinodal decomposition' (Bray 1994).

The demixing kinetics of simple liquids and solids is a well-established field of condensed matter physics in its own right. In metallic alloys, for example, there is a reasonable empirical knowledge of how the process affects the resulting microstructure. Despite extensive investigations, a similar understanding of
`process pathways' in soft condensed matter remains lacking, although important progress was made in the late 1990s, at least of one important class of systems (colloid-polymer mixtures); this is reviewed by Anderson and Lekkerkerker (2002). For example, a colloid  or paste (be it paint, shampoo, or tomato sauce) is
useless if it prematurely separates into a solid block of material with a layer of solvent on top. Whether it will do
this depends on its microstructure; and that in turn depends on its process
history. 

\begin{figure}
\centfig{12cm}{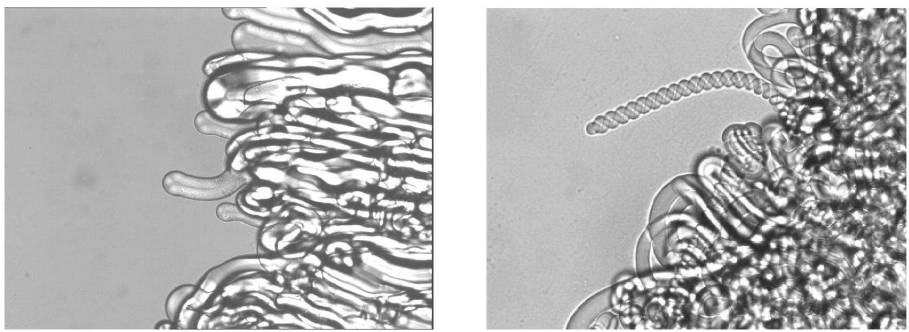}
\caption{The myelinic instability of a dissolving lamellar phase. In both
pictures, the lamellar
phase is on the right; water is on the left. As the two phases mix, the
interface broadens and
myelinic figures are seen. These are long tubular fingers formed from
concentric cylindrical
bilayers. As time passes, the myelins can become more exotic in form, such
as the double helix
shown on the right. (Figure courtesy
of Mark Buchanan.)
\label{fig:myelin}}
\end{figure}

Equally interesting, and less well-studied in other branches of condensed matter physics, is the reverse process. How
do initially separated phases mix, if thermodynamic conditions are changed
to favour this?
For example, a bilayer liquid crystal, as it
dissolves into excess
water, can show various instabilities.  An example is the `myelinic'
instability in
Fig.\ref{fig:myelin}, named after the myelin sheaths which surround nerve
cells in the body,  and
which the observed finger-like patterns closely resemble. This has still
not  been systematically
studied or explained, despite having been first reported in the mid 19th century
(Virchow 1854). One can earnestly hope
for quicker progress in the next century.

\section*{Relation To Other Disciplines}
Soft condensed matter physics forms part of few undergraduate curricula,
and remains
a minority (though a growing) interest among professional condensed-matter
physicists. Among
the latter, it has an undeserved reputation for being a `messy' subject,
meaning perhaps that, in
many soft materials, chemistry as well as physics must be used before a
complete
understanding is obtained. This is true, but the physical themes of large
thermal fluctuations,
constrained Brownian motion, and entropic elasticity recur in system after
system. These provide
a unifying conceptual framework within which  more specific chemical
features can then be
addressed. Although materials scientists, chemists and chemical engineers have had a strong role to play, the recognition of such underlying universalities, and the building of the aforementioned framework linking experiment firmly into statistical mechanics, was arguably the distinctive contribution of physicists to the study of soft condensed matter during the second half of the 20th century.

What about biology? The 1980s and 1990s saw an increasing dialogue between soft matter
physicists and biologists (Lipowsky and Sackmann 1995). One difference between
soft-matter physics and biology is inescapable, however. Physicists are
trained to consider the
generic features of any problem, and then apply the resulting insights to
other situations that are
broadly similar but different in detail. But biological structures have
been honed by evolution so
that the physical exception is (often) the biological norm. In such circumstances, existing
conceptual tools will need to be modified and new ones invented; this vista for future research extends
far into the
21st century.

\section*{Soft Matter and Science Politics}
The physics of soft condensed matter has posed (and continues to pose) stimulating challenges to
experimentalists, theorists, and computer simulators alike. It is all the
more appealing because
it addresses the properties of materials that we encounter every day:
materials we eat, rub on our
skin, decorate our homes with, and (to some extent) are ourselves made
from. The intellectual and
scientific curiosity that this arouses has, I hope,  been partly conveyed
in this historical survey.
Moreover, a deeper understanding of these
properties, and how to control them, can yield important rewards.

The study of soft condensed matter addresses familiar materials; and
familiarity, it is said, breeds contempt. Perhaps this is why, at least during the 1980s and 1990s, the
non-scientific
press often seemed more interested in generating humorous but inaccurate
headlines (`mad scientists
spend government cash shaking tomato sauce') than conveying to their
readers any idea of what the research entailed. A disparaging attitide among colleagues in more `mainstream' areas of research has also be detectable at many times during that period.
In contrast, members of the public have shown delight in learning
that professional scientists (as well as themselves) still do not understand why the tomato
sauce sometimes just won't
come out of the bottle. And few of them doubt that solving this small
problem, alongside a thousand
larger ones, could make their lives in the 21st century a little easier.

\section*{References}

PJ Flory 1953: Principles of Polymer Chemistry, Cornell University Press, Ithaca New York.

PG De Gennes 1979: Scaling Concepts in Polymer Physics, Cornell University Press, Ithaca New York.

M Doi and SF Edwards 1986: The Theory of Polymer Dynamics, Clarendon Press, Oxford.

PN Pusey 1991: Colloidal Suspensions, pp 767--942 in 'Liquids, Freezing and Glass Transition', Proc. Les Houches Session L1, JP Hansen, D Levesque and J Zinn-Justin, eds., Elsevier Science Publishers BV (1991).

VJ Anderson and HNW Lekkerkerker 2002: Insights into phase transition kinetics from colloid science, Nature 416, 811-815.

HNW Lekkerkerker 2000: Phase separation and aggregation in colloidal suspensions, in Soft and fragile matter, nonequilibrium dynamics, metastability and flow, M. E. Cates and M. R. Evans, Eds., IOP Publishing, Bristol.

J Israelachvili 1985: Intermolecular and surface forces, Academic, New York.

RG Larson 1999: The structure and rheology of complex fluids, Oxford University Press. 

ME Cates and SJ Candau 1990: Statics and dynamics of wormlike surfactant micelles, Journal of Physics Condensed Matter 2, 6869-6892.

PG de Gennes 1972: The physics of liquid crystals, Clarendon, Oxford.

G Gompper and M Schick 1994: Self-assembling amphiphilic systems, Academic Press, New York.

R Lipowsky and E Sackmann, Eds, 1995: Structure and dynamics of membranes: From cells to vesicles, Elsevier.

T Vicsek 1989: Fractal Growth Phenomena, World Scientific, Singapore.

MD Haw and WCK Poon 1997: Mesoscopic structure formation in colloidal aggregation and gelation, Advances in Colloid and Interface Science 73, 71--126.

H Rehage and H Hoffmann 1991: Viscoeelastic surfactant solutions - Model systems for rheological research, Molecular Physics 74, 933-973 (1991).

O Diat, D Roux and F Nallet 1992: Effect of shear on a lyotropic lamellar phase, Journal de Physique II 3, 1427-1452.

ME Cates and MR Evans, Eds, 2000: Soft and fragile matter: Nonequilibrium dynamics, metastability and flow, IOP Publishing, Bristol.

AJ Bray 1994: Theory of Phase Ordering Kinetics, Advances in Physics 43, 357-459.

\end{document}